# Effect of Zn substitution on the suppression of $T_c$ of $Y_{1-x}Ca_xBa_2(Cu_{1-y}Zn_y)_3O_{7-\delta}$ superconductors: pseudogap and the systematic shift of the optimum hole content

S. H. Naqib[*a]

[*]*Department of Physics, Rajshahi University, Rajshahi-6205, Bangladesh*
[a]*Shoenberg Laboratory for Quantum Matter, University of Cambridge, J. J. Thomson Avenue, Cambridge CB3 0HE, United Kingdom*

**Abstract**

The effect of Zn substitution on the superconducting transition temperature, $T_c$, was investigated for the sintered $Y_{1-x}Ca_xBa_2(Cu_{1-y}Zn_y)_3O_{7-\delta}$ compounds over a wide range of hole concentration per $CuO_2$ plane, $p$. Ca substitution enabled us to study the deeply overdoped region. $p$ was changed by changing the oxygen deficiency ($\delta$). A strongly $p$-dependent rate of suppression of $T_c$ with Zn ($dT_c/dy$) was found. From the analysis of the $dT_c(p)/dy$ and $T_c(p,y)$ data, we found that the optimum hole content, $p_{opt}$, shifts to higher values with increasing Zn and superconductivity is at its strongest when $p = 0.185 \pm 0.005$. Various complementary experiments have identified this as the hole content where the pseudogap vanishes quite abruptly. We have discussed the possible relevance to these ideas with our findings.

---

[*] Present address. Fax: +88-0721-750064; E-mail: salehnaqib@yahoo.com

# 1 Introduction

Impurity substitution in the cuprate superconductors has proven to be a powerful experimental probe of the basic electronic properties. $T_c$ suppression, the increase of the residual resistivity, the temperature dependence of the penetration depth, the impurity-induced increase of electronic density of states in the low temperature, and impurity-related local bound states seen in STM are just some of the phenomena that have been investigated through impurity studies, and a wealth of information has been garnered [1]. Zn is possibly the most widely substituted impurity in cuprate superconductors. There are several advantages of using Zn, namely: (a) It mainly substitutes the in-plane Cu(2) sites, thus the effects of planar impurity can be studied, (b) the doping level remains nearly the same when Cu(2) is substituted by Zn, enabling one to look at the effects of disorder at almost the same hole concentration [2,3], and (c) Zn reduces $T_c$ most effectively and therefore, provides with a way to look at the low temperature behavior of various "normal state (NS)" properties as superconductivity is suppressed. Most of the studies on effect of Zn on charge transport and magnetic states of $YBa_2Cu_3O_{7-\delta}$ (Y123) are limited to the range of underdoped (UD) to optimum doping (OPD) level. Very few studies are carried out in the overdoped (OD) regions. This is due to a material specific problem associated with this system, namely, the OD side of the phase diagram cannot be studied. Pure Y123 with full oxygenation ($\delta = 0$) is only slightly OD (p ~ 0.18, whilst superconductivity is expected up to p ~ 0.27). Further overdoping can be achieved by substituting $Y^{3+}$ by $Ca^{2+}$, which adds hole carriers in the $CuO_2$ planes independently from the state of oxygenation of the $CuO_{1-\delta}$ chains. 20%Ca substitution plus full oxygenation raises the planar hole content to p ~ 0.23, significantly higher than 0.18 that can be achieved in pure fully oxygenated Y123 [4-6]. Thus study of $Y_{1-x}Ca_xBa_2(Cu_{1-y}Zn_y)_3O_{7-\delta}$ (Ca-Zn-Y123) in the OD region with planar defects can clarify various issues regarding the electronic phase diagram of the cuprates, and Y123 in particular. In this study we have primarily investigated the rate of suppression of $T_c$ with Zn(y) as a function of p over a wide range of values, starting from the UD to the deeply OD regimes. There have been a number of previous studies on the role of Zn on $T_c$ degradation in Y123 at some particular value of p, often in the UD to OPD regions [7-10]. A detailed analysis of the $T_c(p, y)$ behavior over a wide range of hole concentration, from UD to deeply OD region, has not been reported yet. We have found strongly p-dependent rate of suppression of $T_c$ with Zn. There is a fundamental difference in the UD and the OD regimes, namely the superconducting dome ($T_c(p)$) shrinks asymmetrically upon Zn substitution, indicating a difference in the electronic ground states. This is manifested by a systematic shift in the value of the optimum hole content, $p_{opt}$, towards higher values with increasing Zn. Even though the pseudogap (PG) is among the most widely studied phenomena in the physics of the cuprates, the reason for this is still debated [11,12]. The observed asymmetric shift of the superconducting dome can be important in this regard. Our analysis shows that superconductivity is at its strongest when p ~ 0.185 close to p ~ 0.19, reported as the critical hole concentration, $p_{crit}$, (where the PG vanishes) by several earlier studies [4,11,13-15].

## 2 Experimental samples

Polycrystalline single-phase sintered samples of $Y_{1-x}Ca_xBa_2(Cu_{1-y}Zn_y)_3O_{7-\delta}$ (x = 0.05, 0.10, and 0.20; y = 0.005, 0.01, 0.015, 0.03, 0.04, 0.06, and 0.07) were synthesized by standard solid state reaction method [4-6]. Samples were characterized by X-ray diffraction (XRD), electron probe microanalysis (EPMA), ac susceptibility (ACS), resistivity ($\rho(T)$), and room-temperature thermopower (*S[290K]*) measurements. The hole content was varied either by changing the amount of Ca(x) or the oxygen deficiency ($\delta$) by quenching the samples into liquid nitrogen from different temperatures and oxygen partial pressures. Weight changes were noted after each annealing and changes in $\delta$-values were obtained. Details of sample preparation, oxygen annealing, and characterizations can be found in refs. [4,6,15].

## 3 Experimental results

The superconducting transition temperature, $T_c$, was determined from the $\rho(T)$ measurements. $T_c$ was defined at zero resistance (within the instrumental noise level of $\pm\ 10^{-6}\Omega$) (see figure 1a) and the transition width as the temperature difference between 90% and 10% of normal state resistivity, $\rho_n$, defined above superconducting transition. We have also obtained $T_c$ from the low field ($H_{rms}$ = 0.1 Oe; $f$ = 333.3 Hz) ACS measurements as follows: a line was drawn on the steepest part of the diamagnetic ACS curve and another one was drawn as the T-independent base line associated with negligibly small NS signal as shown in figure 1b. The intercept of the two lines gave $T_c$. $T_c$ values obtained by these two methods agree quite well within $\pm$ 1K for all the samples. The transition widths obtained from low-field ACS were always lower than those found from resistivity measurements. This is probably due to the percolative nature of resistive transition, which is greatly affected by the presence of the grain boundaries and voids inside the sample. Low-field ACS, on the other hand, is relatively insensitive to the nature and quality of the grain boundaries. Because of the strong p-dependence of various NS and SC properties (especially near $p_{crit}$) it is important to determine p as accurately as possible. We have estimated p from the room temperature thermopower, *S[290K]* using the correlation of Obertelli *et al* [16]. A previous study showed that *S[290K]* does not change much with Zn content in Zn-substituted Y123 for fixed values of $\delta < 0.5$ [3]. We found the same trend in $Y_{1-x}Ca_xBa_2(Cu_{1-y}Zn_y)_3O_{7-\delta}$ [4,6,15,17]. Accordingly, in this doping range *S[290K]* is still a good measure of p even in the presence of strong in-plane scattering by Zn. For Zn-free samples we have also estimated p using the parabolic $T_c(p)$ relation [18], given by

$$T_c(p) = T_{max}(1 - Z(p - p_{opt})^2) \qquad (1)$$

with the usual values $Z = 82.6$ and $p_{opt} = 0.16$ (for maximum $T_c$) [18]. Using p from *S[290K]* and the experimental values of $T_c$, we have obtained a very good fit of $T_c(p)$ with equation (1) for all $Y_{1-x}Ca_xBa_2(Cu_{1-y}Zn_y)_3O_{7-\delta}$ samples. Representative plot for the $Y_{0.80}Ca_{0.20}Ba_2(Cu_{1-y}Zn_y)_3O_{7-\delta}$ compounds are shown in figure 2. The values of $T_{cmax}$, Z, and $p_{opt}$, respectively, are 0% Zn: 84.5K, 82.6, and 0.16; 1.5% Zn: 65.5K, 146, and 0.162; 3% Zn: 51.8K, 205, and 0.165; and 6% Zn: 27.5K, 410, and 0.174. This systematic shift of $p_{opt}$ towards higher values with increasing Zn (y) bears particular significance. Our analysis of the $dT_c(p)/dy$ data indicates that superconductivity is at its strongest at p ~ 0.185 (i.e., the largest amount of Zn is needed to suppress superconductivity completely at this hole concentration). This is in complete agreement with earlier studies [11,13,19] where superconducting condensation energy and superfluid density were found to be maximum at p ~ 0.19 for various different families of cuprates. The systematic variations of $p_{opt}(y)$ and $Z(y)$ were almost identical for all the samples, independent of Ca content. This implies that in-plane (Zn) and out-of-plane (Ca) disorders affect the various SC and NS properties in fundamentally different fashion. The most obvious and striking effect of Zn substitution in cuprates is the extremely rapid suppression of $T_c$. The rate of suppression of $T_c(p)$ with Zn, $dT_c/dy$, in $Y_{1-x}Ca_xBa_2(Cu_{1-y}Zn_y)_3O_{7-\delta}$), depends strongly on the planar hole content p. Figure 3 shows $T_c(p)$ versus y for some of the $Y_{0.80}Ca_{0.20}Ba_2(Cu_{1-y}Zn_y)_3O_{7-\delta}$ compounds. We have defined $y_c(p)$ [$\equiv T_c(p; y = 0)/(dT_c(p)/dy)$] as the critical Zn concentration needed to suppress superconductivity completely for a given p. The behavior of $y_c(p)$ is shown in figure 4. The variation of the fitting parameter Z (which determines the width of the SC dome) with Zn concentration is shown in figure 5a. $p_{opt}$ versus y is plotted in figure 5b. Tables 1 and 2 present the values of $Z(y)$, $p_{opt}(y)$, $T_{cmax}(y, p_{opt})$ and $dT_c(p)/dy$, $y_c(p)$, respectively for all the $Y_{1-x}Ca_xBa_2(Cu_{1-y}Zn_y)_3O_{7-\delta}$ compounds under study. The variation of $T_{cmax}(y)$ versus $p_{opt}(y)$ is shown in figure 6 for the $Y_{0.80}Ca_{0.20}Ba_2(Cu_{1-y}Zn_y)_3O_{7-\delta}$ compounds. Figure 6 agrees with the rough estimates shown in figure 2 in the sense that $p_{opt}(y) \sim 0.185$ as $T_{cmax}(y)$ tends to zero.

## 4 Discussions

Figure 5b shows clearly the shift in $p_{opt}(y)$ and therefore the parabolic $T_c(p)$ dome. Qualitatively this shift might be attributed due to the existence of the PG in the quasi-particle spectral density. Regardless of the origin of the PG, the primary effect of its presence is the removal of the low-energy quasi-particle states which otherwise would have been available for superconducting condensation of the Cooper pairs. Zn acts as unitary scatterer and breaks the paired carriers most effectively [20]. The effect of Zn is more pronounced when the PG is present. This is mainly due to the facts that the SC order parameter has d-wave symmetry and in the unitary limit the pair-breaking scattering rate is inversely proportional to the electronic density of states near the chemical potential. Since both PG (which grows as p decreases) and Zn reduces the superfluid density (Zn and Ca does not affect the magnitude of the PG) it is natural that in the Zn substituted compounds the (*yet to be understood*) condition for maximum $T_c$ is satisfied at a higher hole content (giving the *appropriate* superfluid density and SC condensation energy) compared to the pure compounds. The asymmetrical shrinkage of the SC dome also implies that the onset of superconductivity in the impurity substituted UD region is greatly affected by the presence of the PG. The OD region of the T-p phase diagram of the cuprate superconductors are widely believed to be relatively better understood. But answers to some very fundamental questions are yet to be found. For example, the reason why $T_c$ decreases in the OD side? $y_c(p)$ in the OD region, where $p > 0.185$, grossly follows the $T_c(p)$ behavior as the pair-breaking scattering rate becomes almost constant. The increase of $dT_c(p)/dy$, though significantly lower compared to that in the UD region, in the overdoped side for $p > 0.185$ is hard to explain. For one thing, the low-energy electronic density of states stays almost the same for $p > 0.185$ in $Y_{1-x}Ca_xBa_2(Cu_{1-y}Zn_y)_3O_{7-\delta}$ [6,14], which should give a nearly constant pair-breaking scattering rate in the unitary limit. One possible reason for the increase in $dT_c(p)/dy$ could be due to the sharp reduction in the superfluid density in the overdoped region. In principle, same impurity scattering is expected to have different effects on pair-breaking depending on the strength of superconductivity. A simple Abrikosov-Gor'kov [20,21] type analysis does not include this in the formalism and a more realistic model is required to explain $dT_c(p)/dy$ over the whole doping range. The shrinkage of the SC dome in the OD side is thus probably due to the rapid reduction of the superfluid density due to some correlations (e.g., separation into non-SC metallic and SC phases [22]) that are detrimental to superconductivity.

## 5 Conclusions

We have presented a detailed analysis of the effect of Zn on $T_c$ of $Y_{1-x}Ca_xBa_2(Cu_{1-y}Zn_y)_3O_{7-\delta}$ compounds over a wide range of hole concentration. From the analysis of the $T_c(p, y)$ data we have shown how the width of the SC dome and the hole content for maximum $T_c$, $p_{opt}$, vary with Zn. The shift of $p_{opt}(y)$ towards higher values and thus the asymmetry in the superconducting dome has been attributed to the presence of the PG for compounds with $p < 0.185$. The evolution of both $dT_c(p)/dy$ and $y_c(p)$ imply that superconductivity is at its strongest at $p \sim 0.185$, where the PG vanishes. The findings in this study are consistent with the scenario in which PG competes (at least in the sense that it removes the quasi-particle states which otherwise would have been available for SC condensation) and coexists with superconductivity. In the overdoped side $dT_c(p)/dy$ shows slowly increasing trend with increasing hole concentration. The reason for this behavior is not entirely clear to us and a better theoretical understanding is required to resolve this issue.

**Acknowledgements**

The author thanks Dr. J. R. Cooper and Dr. J. W. Loram of University of Cambridge, UK, and Prof. J. L. Tallon of MacDiarmid Institute for Advanced Materials and Nanotechnology and Victoria University at Wellington, New Zealand, for their helpful comments and suggestions at various stages. The author would also like to thank the Commonwealth Scholarship Commission, UK, and IRC in Superconductivity, University of Cambridge, UK, for funding this research.

**Table 1**

$Z(y)$, $p_{opt}(y)$ (obtained from fitting equation (1)), and experimental $T_{cmax}$ for $Y_{1-x}Ca_xBa_2(Cu_{1-y}Zn_y)_3O_{7-\delta}$ compounds.

| x | y | Z | $p_{opt}$ | $T_{cmax}$ (K) ($\pm 1.0$) |
|---|---|---|---|---|
| 0.05 | 0.00 | 82.6 | 0.160 | 91.00 |
| | 0.02 | 168.0 | 0.165 | 70.40 |
| | 0.04 | 281.0 | 0.169 | 52.10 |
| | 0.07 | 521.3 | 0.178 | 23.90 |
| 0.10 | 0.00 | 82.6 | 0.160 | 88.00 |
| | 0.015 | 133.7 | 0.162 | 68.22 |
| | 0.03 | 208.0 | 0.164 | 52.06 |
| 0.20 | 0.00 | 82.6 | 0.160 | 84.50 |
| | 0.015 | 146.0 | 0.163 | 65.50 |
| | 0.03 | 205.0 | 0.165 | 51.80 |
| | 0.04 | 280.0 | 0.168 | 44.20 |
| | 0.06 | 410.0 | 0.174 | 27.50 |

**Table 2**

$dT_c(p)/dy$ and $y_c(p)$ for $Y_{1-x}Ca_xBa_2(Cu_{1-y}Zn_y)_3O_{7-\delta}$.

| Ca content (x) | p (± 0.004) | $dT_c/dy$ (K/%Zn) (± 0.20) | $y_c$ (% of Zn) (± 0.10) |
|---|---|---|---|
| 0.20 | 0.202 | -9.50 | 7.57 |
| | 0.193 | -8.70 | 8.84 |
| | 0.184 | -8.80 | 9.14 |
| | 0.170 | -9.32 | 8.80 |
| | 0.161 | -9.80 | 8.60 |
| | 0.145 | -10.60 | 7.80 |
| | 0.130 | -11.70 | 6.68 |
| | 0.121 | -12.29 | 6.01 |
| | 0.110 | -14.20 | 4.72 |
| 0.10 | 0.204 | -9.90 | 7.57 |

|  | 0.195 | -8.40 | 9.55 |
|  | 0.186 | -8.00 | 10.54 |
|  | 0.170 | -9.40 | 9.40 |
|  | 0.165 | -10.00 | 8.91 |
|  | 0.123 | -12.40 | 6.38 |
|  | 0.104 | -15.34 | 4.31 |
| 0.05 | 0.189 | -8.20 | 10.39 |
|  | 0.174 | -9.20 | 9.80 |
|  | 0.154 | -10.31 | 8.86 |
|  | 0.136 | -11.07 | 7.88 |
|  | 0.116 | -12.90 | 6.00 |

Figure 1. (Color online) Determination of $T_c$ from (a) resistivity and (b) ac susceptibility measurements for two representative $Y_{1-x}Ca_xBa_2(Cu_{1-y}Zn_y)_3O_{7-\delta}$ compounds. Ca(x) and Zn(y) contents are shown.

Figure 2. (Color online) $T_c$ versus p for $Y_{0.80}Ca_{0.20}Ba_2(Cu_{1-y}Zn_y)_3O_{7-\delta}$. y-values are shown in the plot. The dashed lines are $T_c(p)$ fits of equation (1), the full straight line shows roughly the trend of $p_{opt}(y)$.

Figure 3. (Color online) $T_c$ versus Zn concentration (y) for $Y_{0.80}Ca_{0.20}Ba_2(Cu_{1-y}Zn_y)_3O_{7-\delta}$ compounds with different values of p. p-values are given in the figure.

Figure 4. (Color online) The critical Zn concentration, $y_c$, versus p for $Y_{1-x}Ca_xBa_2(Cu_{1-y}Zn_y)_3O_{7-\delta}$. The gray vertical lines locate the hole contents 0.16 (optimum p for Zn-free compound) and 0.185 (the critical hole content where superconductivity is at its strongest).

Figure 5. (Color online) (a) Z versus y and (b) $p_{opt}$ versus y for $Y_{1-x}Ca_xBa_2(Cu_{1-y}Zn_y)_3O_{7-\delta}$ samples (see text for details). Ca contents (x) are given in the figure. Dashed lines are guide for the eyes.

Figure 6. (Color online) Maximum superconducting transition temperature, $T_{cmax}(y)$ versus $p_{opt}(y)$ for $Y_{0.80}Ca_{0.20}Ba_2(Cu_{1-y}Zn_y)_3O_{7-\delta}$ compounds. Dashed line is a guide to the eyes.

Figure 1.

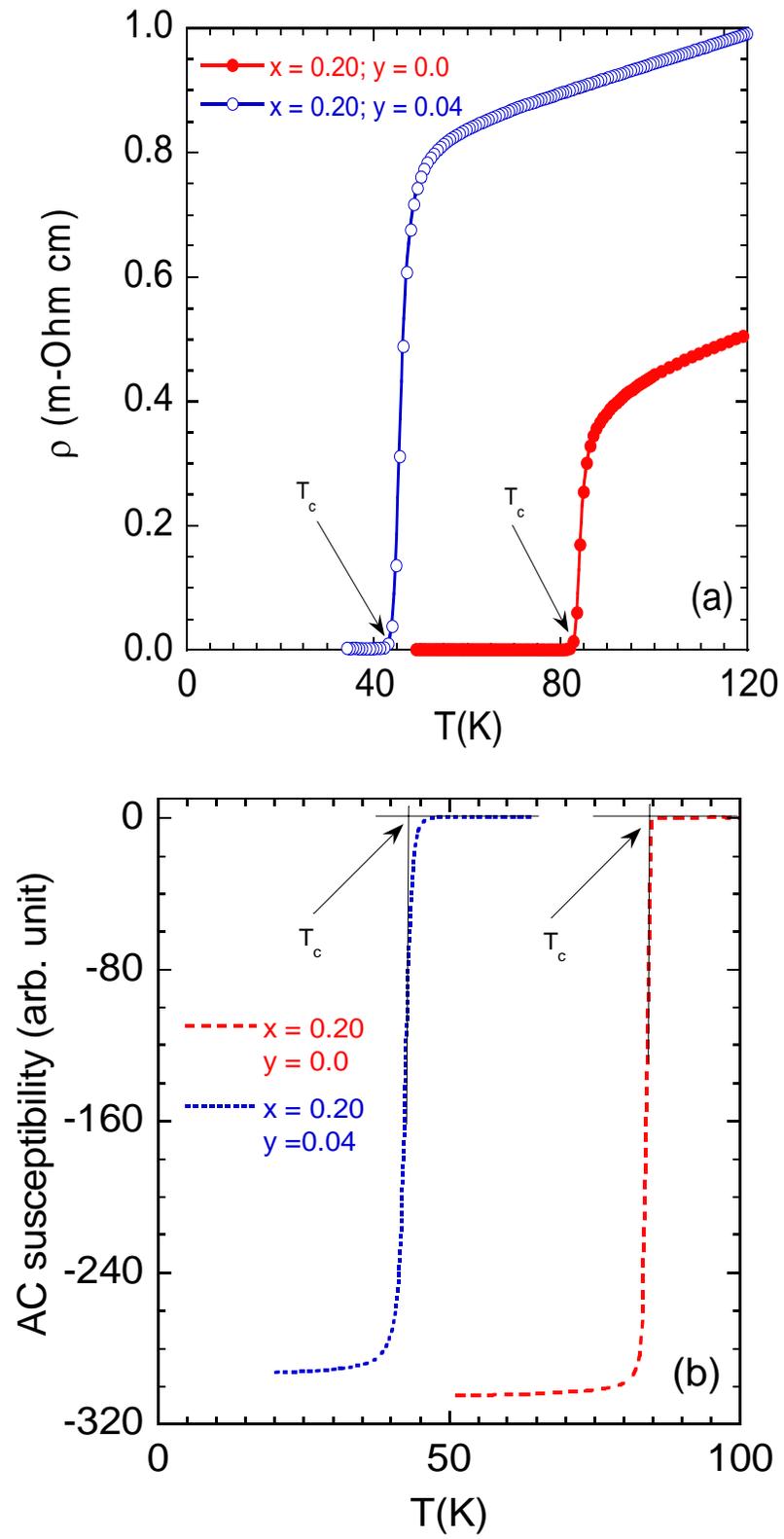

Figure 2.

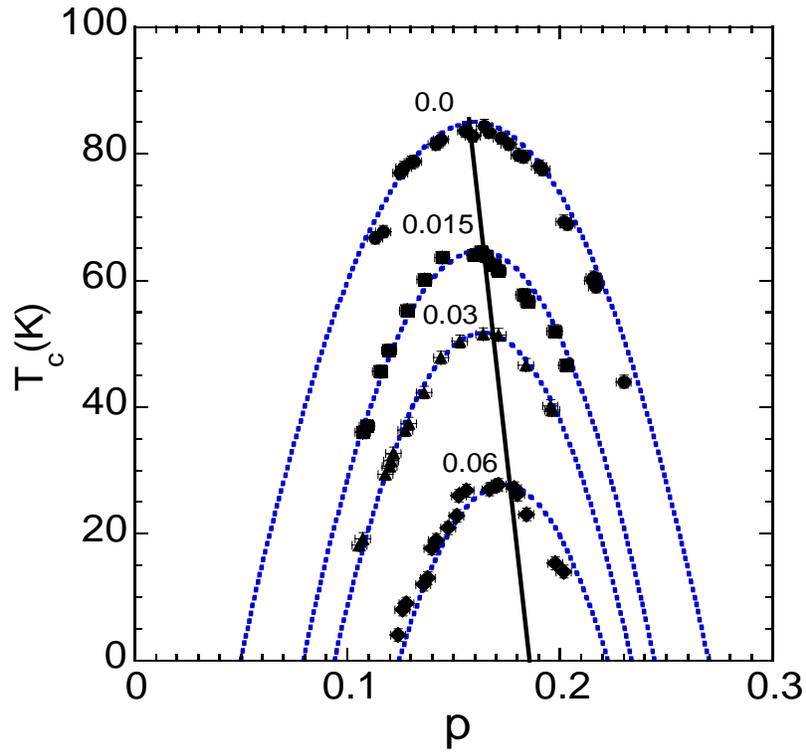

Figure 3.

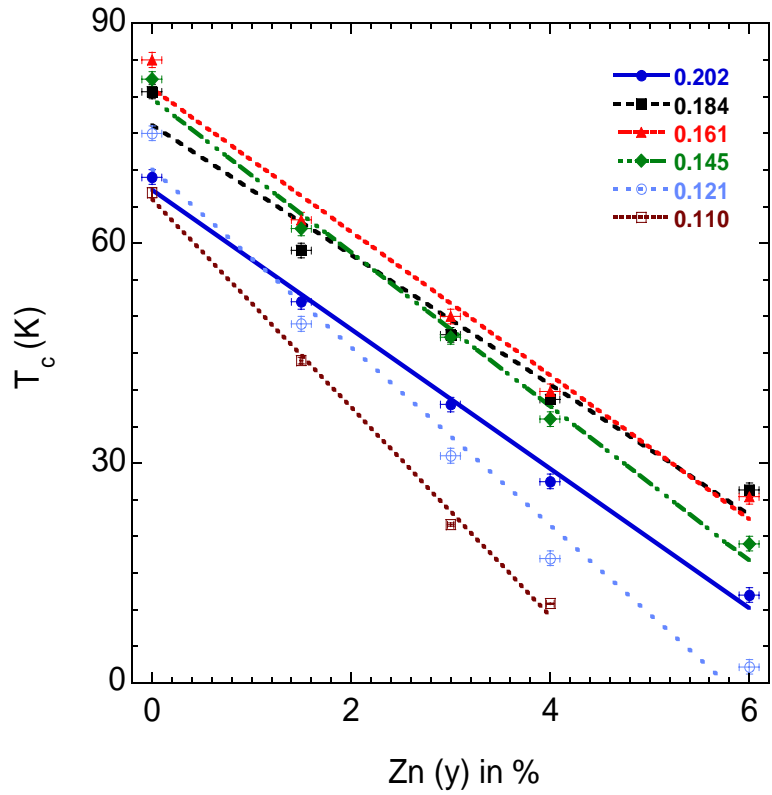

Figure 4.

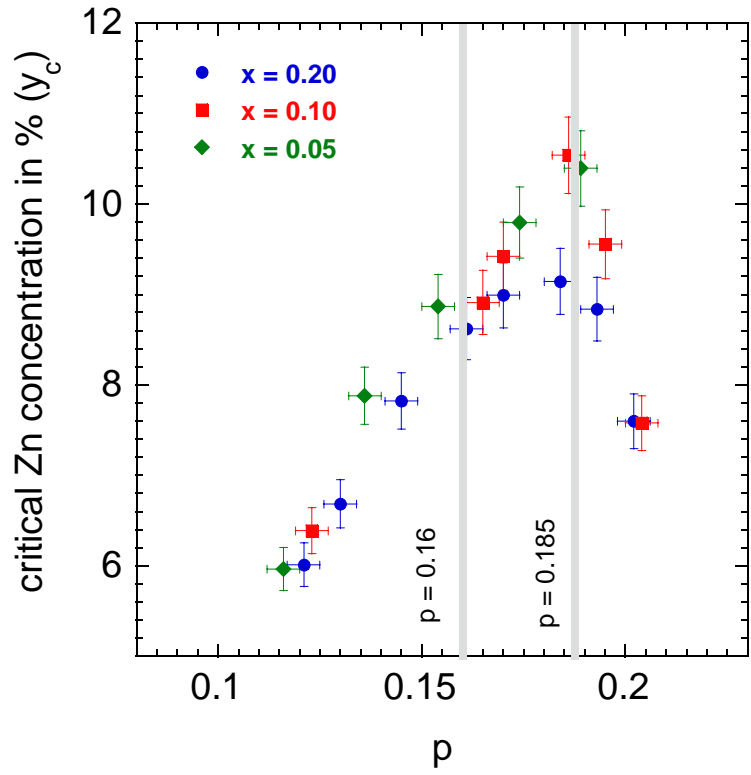

Figure 5.

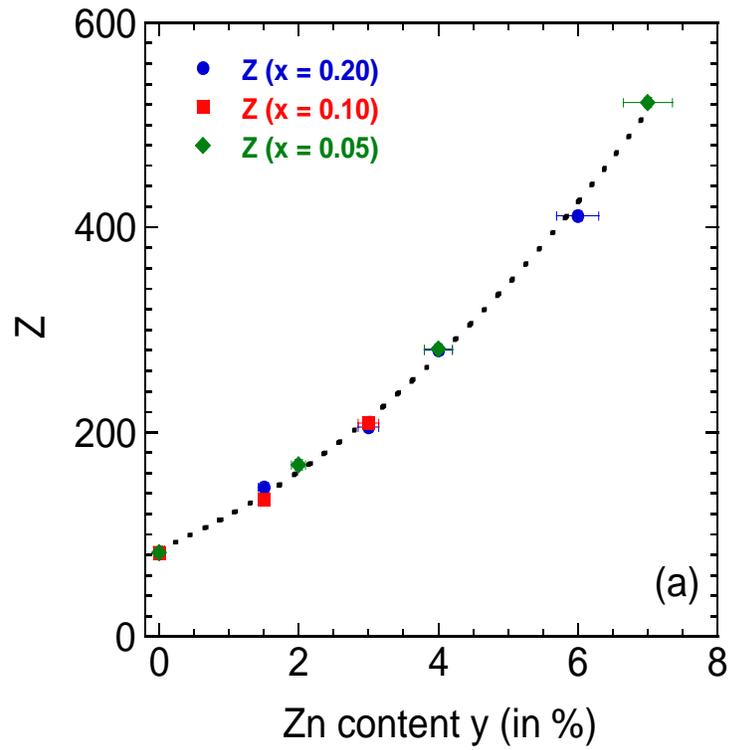

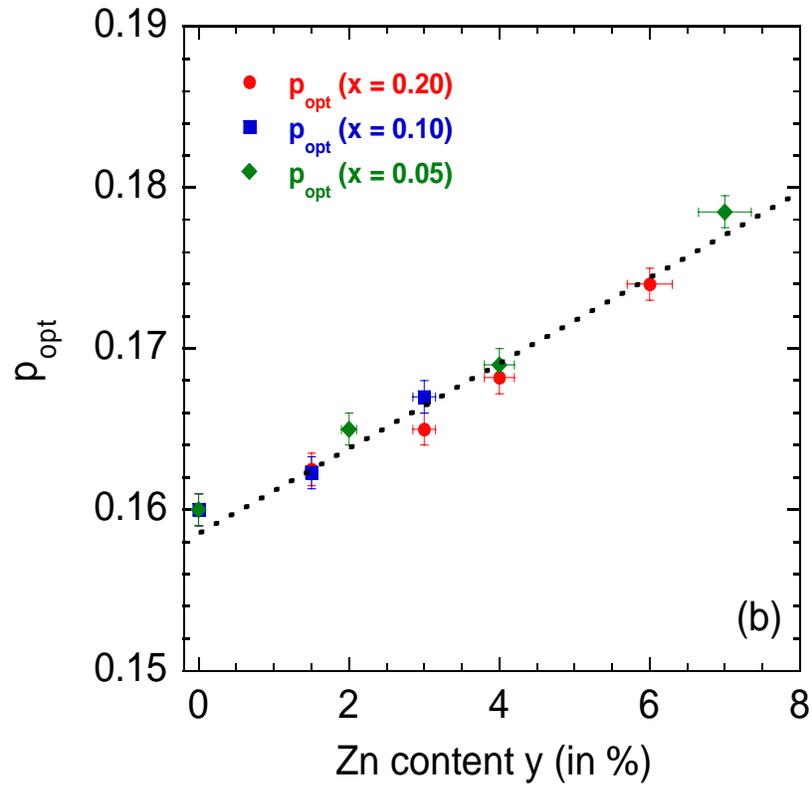

Figure 6.

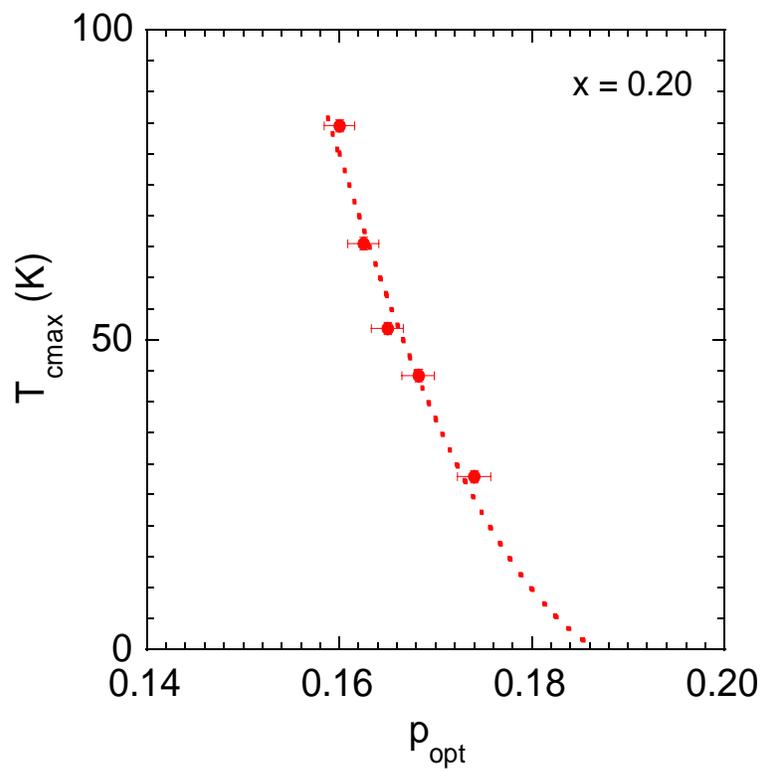